\documentclass[12pt, letterpaper]{article}
\usepackage[top=1in, bottom=1.5in, left=1in, right=1in]{geometry}
\usepackage[utf8]{inputenc}
\usepackage{booktabs}
\usepackage{hyperref}
\usepackage{graphicx}

\title{Deep Drilling Fields for Solar System Science}
\author{David E. Trilling, Michele Bannister, Cesar Fuentes, 
David Gerdes, \\
 Michael Mommert, Megan E. Schwamb, Chad Trujillo}
\date{30 November 2018}

\begin{document}

\maketitle

\begin{abstract}
We propose an ecliptic Deep Drilling Field that will 
discover some 10,000~small and faint Kuiper Belt Objects (KBOs) --- primitive rocky/icy bodies that orbit at the outside
of our Solar System and uniquely record the processes of planetary system
formation and evolution.
The primary goals are to measure the KBO size and shape distributions down
to 25~km, a size that probes both the early and ongoing evolution of
this population.
These goals can be met with around 10~hours total of on-sky time (five separate fields
that are observed for 2.1~hours each).
Additional science will result from downstream observations that provide
colors and orbit refinement, for a total time request of 40~hours over
the ten year LSST main survey.
\end{abstract}

\section{White Paper Information}

\begin{enumerate} 
\item {\bf Contact Information:} David E. Trilling ({\tt david.trilling@nau.edu})
\item {\bf Science Category:} which of the four main LSST science themes are addressed? Are there other
science programs addressed by this white paper?
\item {\bf Survey Type Category:} Deep Drilling Field
\item {\bf Observing Strategy Category:} an integrated program with science that hinges on the combination of pointing and detailed
observing strategy
\end{enumerate}  

\clearpage

\section{Scientific Motivation}


\vspace{1ex}

\noindent The space beyond the orbit of Neptune is inhabited by planetesimals left over from our Solar System's formation 4.5 billion years ago. These Kuiper Belt Objects (KBOs) preserve many clues 
about the evolution of our planetary system.
KBOs shine in reflected light, which makes observing small and distant KBOs very challenging. As a result, our knowledge of the trans-Neptunian region is based predominantly on the study of only the brightest objects (diameters diameters 100--300 km [1]). Our understanding of the smallest (and most numerous) KBOs is generally driven by theoretical models [2], indirect evidence [3, 4, 5], and analysis of the sparse existing data [6, 7].

\vspace{1ex}

\noindent Bernstein et al.\ [8] established the state of art in faint KBO searches by surveying 0.02~deg$^2$ with HST's ACS/WFC,
using more than 100 orbits to discover three new objects to a limiting magnitude of R=28.5. 
Critically, Bernstein used a shift-and-stack technique to detect KBOs fainter than the noise threshold in individual images.
Fuentes et al.\ [9, 10] placed further constraints on the small KBO population by using archival HST data to detect objects smaller than the break in the size distribution. 

\vspace{1ex}

\noindent The observed break in the KBO size distribution (Figure~1)
is generally attributed to collisional evolution over the age of the
Solar System, but may alternatively result from the early process of
planetesimal formation.
If the collisional scenario is correct, objects with sizes smaller than the break radius will have had their surface properties modified, and should therefore display different colors than larger objects. Collisional resurfacing is thought to expose bluer (fresher) material, suggesting that the typical color of smaller objects should be bluer than their larger counterparts. This collisional evolution also sheds light on the processes that govern debris disks. The outstanding question is therefore ``Is there currently collisional
evolution occurring in the distant outer Solar System?'' 
This question has broad implications, from understanding 
the evolution of our Solar System to interpreting the presence of interstellar interloper `Oumuamua in the Solar
System and unraveling the geophysics of Pluto based on interpreting its crater record.

\vspace{1ex}

\noindent Unraveling these mysteries in the outer Solar System requires a large catalog of small, faint KBOs, which in 
turn requires a large telescope with a large field of view. LSST will be the premier deep KBO search facility. 
Many significant questions can be addressed with a small amount of time in a dedicated Deep Drilling Field (DDF).

\vspace{1ex}

\noindent {\em We propose here an ecliptic DDF program that will revolutionize our understanding of the outer Solar System.
This program consists of visits to five different fields. 
Each visit requires 2.1~hours of continuous observing of a single pointing, which gives
a stacked imaged depth of around $r=27.5$ (5$\sigma$), which corresponds roughly
to KBOs of diameter around 25~km -- well below the break in the size distribution.
The expected yield is around $10^4$~KBOs, which gives a fractional error of around
1\% on the cumulative luminosity function of very small KBOs.
The total time request is 10.5~hours for the primary science goals and
42~hours to meet a number of additional science goals.}

\vspace{1ex}

\noindent With
just a single 2.1~hour visit to each of
the five fields the following science goals can be met:
(1) Measuring the {\em KBO size distribution down to $\sim$25 km}, where it is very poorly
known through an exceedingly small sample size; and
(2) Measuring the {\em shape distribution of KBOs} from (partial) lightcurves for objects brighter than 10$\sigma$.

\vspace{1ex}

\noindent Additional science can be provided with three additional visits --- one off-opposition (nominally in the same year as the initial
observation), and one more in each of two other years. 
These four additional visits allow clear dynamical classification of the detected objects through having a two
year arc (from first observation to fourth observation) that includes an off-opposition measurement.
Furthermore,
at least one of these should be in a different filter.
Thus, both orbit refinement and color information is available. With a total of four visits to each field,
as described,
the following science goals can be met:
(3) Measuring the {\em colors of thousands of faint KBOs}; and
(4) Measuring colors, size distribution, and shapes as a {\em function of dynamical class and of size}.
In both cases, differences across 
subgroups constrain
the dynamical evolution of the early Solar System.

\vspace{1ex}

\noindent With this survey we will increase the number of 50~km KBOs by {\em a factor of~40}. 
Finally, at the conclusion of this program our objects
will have
well-enough known orbits that they can be targeted by JWST for further in-depth analysis (i.e., spectroscopy).
This survey will be more than ten times larger than the
largest KBO survey to date (the OSSOS survey, which has discovered some 800~KBOs; see [14] and many other papers).

\vspace{1ex}

\noindent There are a number of ancillary science goals that can be met with the proposed data set:
Measuring the size distribution and colors of Centaurs --- escaped KBOs and progenitors of comets --- to a few kilometers in size;
Measuring the colors of thousands of main belt asteroids (as a function of size);
Measuring the shape distributions and rotational properties of thousands of main belt asteroids (as a function of size);
Possibly discovering and measuring the sizes, colors, and shapes of outer planet Trojan asteroids;
Searches for objects in unusual orbits (including retrograde orbits);
Searches for activity in faint (small) KBOs, Centaurs, and asteroids;
Providing interesting targets that can be studied in detail with JWST;
Searches for distant planets (i.e., planet X) in the far outer Solar System;
Searches for supernovae or other transient events in deep image
pointings.
Some of these ancillary science goal can be met with just the single
first visit; others require the three additional visits as well.

\vspace{1ex}

\noindent This is an ideal project to carry out with LSST. DECam does not go as deep and does not have as wide
a field of view. HSC also has a much smaller field of view, making this project very difficult to carry out
at Subaru. Details of the required sample size are given in the technical sections below.

\vspace{1ex}

\noindent There are already four approved DDFs, all with primary science that is cosmology-oriented. None of these
fields is on or close to the ecliptic, so those data sets will not contain KBOs on ``primordial'' (low inclination) orbits.
Thus, no ``existing'' LSST observations meet the science requirements of this program, although we do note that the
off-ecliptic DDFs may provide useful comparisons between the low- and high-inclination populations.

\clearpage

\vspace{-10ex}

\begin{figure}
\begin{center}
\includegraphics[width=9cm,angle=270]{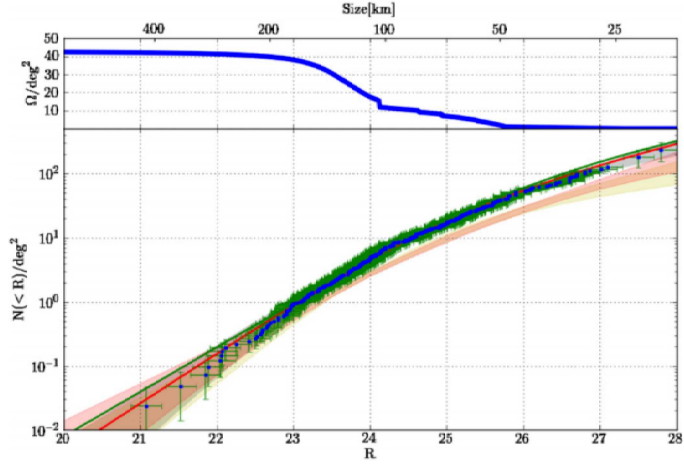}
\end{center}
\vspace{-4ex}
\caption{Figure from [9].
Top panel shows in blue the effective area for all surveys considered in that paper as a function of R magnitude, normalized by the effective area at each
magnitude. The lower panel shows the luminosity function of KBOs, normalized by the effective area at each magnitude. Two different
model fits to the data are shown in green and red.
The gray area corresponds to the 1$\sigma$ confidence
region given for all objects. The same confidence regions are given for hot and cold (high and
low inclination dynamical subclasses) objects, in red and yellow, respectively.}
\label{fuentes}
\end{figure}

\begin{figure}
\begin{center}
\includegraphics[width=12cm]{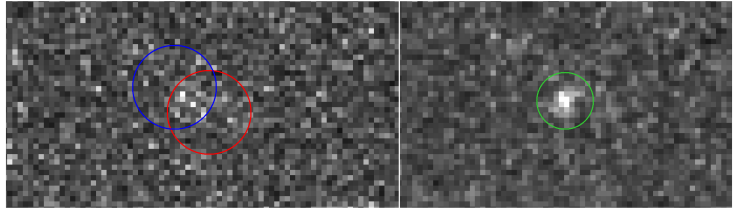}
\end{center}
\vspace{-3ex}
\caption{Illustration of digital tracking for 2013 RF98 ($r$-band magnitude = 24.4; a = 325 AU). \textit{Left}: Image of 2013 RF98 stacked in the sky frame consisting of  11$\times$330-second $z$-band exposures (processed by the DES difference imaging pipeline) in a DES deep supernova field. The blue circle shows the position of 2013 RF98 in the first exposure, and the red circle shows the position of 2013 RF98 in the last exposure. No source was detected at this position in this image. \textit{Right}: The same individual exposures as to the left, but shifted and stacked at the speed of 2013 RF98. The object is detected with SNR$\sim$16.}
\label{stack}
\end{figure}

\clearpage

\vspace{.6in}

\section{Technical Description}
\begin{footnotesize}
{\it Describe your survey strategy modifications or proposed observations. Please comment on each observing constraint
below, including the technical motivation behind any constraints. Where relevant, indicate
if the constraint applies to all requested observations or a specific subset. Please note which 
constraints are not relevant or important for your science goals.}
\end{footnotesize}

\subsection{High-level description}
\begin{footnotesize}
{\it Describe or illustrate your ideal sequence of observations.}
\end{footnotesize}

\vspace{1ex}

\noindent Five footprints on the sky are to be observed. The minimum science goal requires just a single continuous (2.1~hour) stare
at each point. The nominal plan uses just a single filter in each continuous stare. There is a marginal science gain
from using the standard WFD-like 30~second exposure, but it is more efficient to use long (e.g., 600~sec) exposures; this is a 
trade that can be studied in more detail.
Enhanced science goals require three additional visits to each of the five fields. The additional visits are
1--2~months later in year~1; in year~2; and in year~3 or~4.
The total time for the minimum science goal is just 10.5~hours; for the additioanl science goals, the total time
needed is around 42~hours.
The five fields have independent placements on the sky but should all be on the ecliptic. This program should be relatively
easy to schedule, given its overall flexibility of pointing, timing, etc.

\vspace{.3in}

\subsection{Footprint -- pointings, regions and/or constraints}
\begin{footnotesize}{\it Describe the specific pointings or general region (RA/Dec, Galactic longitude/latitude or 
Ecliptic longitude/latitude) for the observations. Please describe any additional requirements, especially if there
are no specific constraints on the pointings (e.g. stellar density, galactic dust extinction).}
\end{footnotesize}

\vspace{1ex}

\noindent The five pointing should be on the ecliptic plane (though there is a nuance involving trade-offs for
location(s) on the invariable plane instead; the difference is small). The galactic plane should be avoided.
Pointing(s) that include some parts of the Neptunian Trojan clouds --- areas near RA of 4~hours
and 20~hours in 2025 --- are preferred for additional science. We note that each pointing is advanced each year
by the mean motion of KBOs as seen from Earth in order to maximize the number of objects that stay within
the field.

\vspace{1ex}

\noindent The five different pointings allow us to sample a range of 
ecliptic longitudes and therefore different positions relative to Neptune, which drives
the resonances in the outer Solar System. Ideally, one of the pointings should include a Trojan cloud
of Neptune, where dynamically stable objects orbit in a 1:1~resonance with Neptune.
Since this resonance is known to be stable over the lifetime of the Solar System, 
the physical properties of objects trapped in this resonance strongly constrain models
of Solar System formation.

\subsection{Image quality}
\begin{footnotesize}{\it Constraints on the image quality (seeing).}\end{footnotesize}

\vspace{1ex}

\noindent While all science programs benefit from the best image quality, this program does not specifically
demand that performance. The key here is depth. Since poor image quality reduces point source depth,
we propose that ``typical'' image quality or better (e.g., from {\tt minion\_1016})
would be preferred. This likely implies somewhat low
airmass. Furthermore, image stability over
the hours of staring is critical so that the shift-and-stack method has consistent images
to combine.

\subsection{Individual image depth and/or sky brightness}
\begin{footnotesize}{\it Constraints on the sky brightness in each image and/or individual image depth for point sources.
Please differentiate between motivation for a desired sky brightness or individual image depth (as 
calculated for point sources). Please provide sky brightness or image depth constraints per filter.}
\end{footnotesize}

\vspace{1ex}

\noindent As described below, there is a trade study to be done here. The science can be accomplished {\em either}
with a series of 30~second (WFD-like) exposures or with a series of longer (perhaps 600~second) exposures.
In the former case, by using a WFD-like cadence, other science cases may benefit from this DDF
program. However, the overhead of 2~seconds per frame adds significant cost to the execution of this program.
Furthermore, there are many more frames to be combined through shift-and-stack, and this increases
the computational need.
In the latter case (long exposures), the data is less likely to be useful for other science
cases, but the overhead is significantly less.
The overall science goal is to reach $r=27.5$ (5$\sigma$) and is agnostic to the shorter/longer exposures trades described above.
On balance, the shorter (WFD-like) exposure that provide additional science (for us and for others) are preferred
scientifically, but the efficiency gains from longer exposures are recognized. This trade can be studied in further
detail.

\subsection{Co-added image depth and/or total number of visits}
\begin{footnotesize}{\it  Constraints on the total co-added depth and/or total number of visits.
Please differentiate between motivations for a given co-added depth and total number of visits. 
Please provide desired co-added depth and/or total number of visits per filter, if relevant.}
\end{footnotesize}

\vspace{1ex}

\noindent The required total co-added image depth is $r=27.5$ as motivated in the science section above. 
That image depth must be obtained in each visit (2.1~hours of continuous observing).

\subsection{Number of visits within a night}
\begin{footnotesize}{\it Constraints on the number of exposures (or visits) in a night, especially if considering sequences of visits.  }
\end{footnotesize}

\vspace{1ex}

\noindent As described above, each visit must be carried out as 2.1~hours of continuous imaging.

\subsection{Distribution of visits over time}
\begin{footnotesize}{\it Constraints on the timing of visits --- within a night, between nights, between seasons or
between years (which could be relevant for rolling cadence choices in the WideFastDeep. 
Please describe optimum visit timing as well as acceptable limits on visit timing, and options in
case of missed visits (due to weather, etc.). If this timing should include particular sequences
of filters, please describe.}
\end{footnotesize}

\vspace{1ex}

\noindent For the minimum science case, each field is visited just once, so this question is not relevant.
For the additional science cases, the second visit to each field should be 1--2~months after the first visit;
the third visit should be $\sim$1~year after the first visit; and the fourth visit should be 
$\sim$2~years after the first visit.

\vspace{1ex}

\noindent With just a single year, the primary science goals can be met, but none of the additional
science goals described in the science justification.

\vspace{1ex}

\noindent Our simulations have shown that dynamical classification of KBOs is not really certain until
the orbital elements are known to better than 1\%. In general, for KBOs this requires a two
year arc that includes an off-opposition astrometric measurement. This is the cadence proposed here.

\subsection{Filter choice}
\begin{footnotesize}
{\it Please describe any filter constraints not included above.}
\end{footnotesize}

\vspace{1ex}

\noindent The minimum science requirement is a single filter, presumably $r$, which is the 
most sensitive for the typically reddish Solar System objects.
The additional science goals require that for each pointing one of the additional
visits should be in a second filter, and $g$ probably provides the best compromise
between throughput and diagnostic color [15].
The $g-r$ color implies intrinsic composition and correlations between color and
dynamical subclass, while still somewhat marginal, can constrain the dynamics of the early
Solar System. Increasing the number of faint KBOs with measured colors by 1000~objects (as described below)
will provide significant power on constraining these formation models.

\vspace{1ex}

\noindent Colors for individual objects will be derived by first measuring $H_{band}$, the Solar System absolute
magnitude of an object in a given band, which can be derived from each individual 2.1~hour visit. Thus, the 
$g-r$ (for example) color is found as $H_g - H_r$. We note (as described below) that high-fidelity colors
of course will not be available for objects detected at 5$\sigma$; 30$\sigma$ in each band is a more reasonable requirement
for color determinations. Thus, color measurements will be available for objects brighter than
around~25.5; at 30~KBOs/deg$^2$ this gives 1440~KBOs with colors across five pointings for objects
as small as 50~km. This will be highly diagnostic for understanding the evolution of small
KBOs.

\subsection{Exposure constraints}
\begin{footnotesize}
{\it Describe any constraints on the minimum or maximum exposure time per visit required (or alternatively, saturation limits).
Please comment on any constraints on the number of exposures in a visit.}
\end{footnotesize}

\vspace{1ex}

\noindent As described below, there is a trade study to be done here. The science can be accomplished {\em either}
with a series of 30~second (WFD-like) exposures or with a series of longer (perhaps 600~second) exposures.
In the former case, by using a WFD-like cadence, other science cases may benefit from this DDF
program. However, the overhead of 2~seconds per frame adds significant cost to the execution of this program.
Furthermore, there are many more frames to be combined through shift-and-stack, and this increases
the computational need.
In the latter case (long exposures), the data is less likely to be useful for other science
cases, but the overhead is significantly less.
The overall science goal is to reach $r=27.5$ (5$\sigma$) and is agnostic to the shorter/longer exposures trades described above.

\subsection{Other constraints}
\begin{footnotesize}
{\it Any other constraints.}
\end{footnotesize}

\vspace{1ex}

\noindent This imaging is best carried out in dark conditions. Because the individual continuous stares
are not long, at just 2.1~hours each, they hopefully will not be interrupted by changes in weather,
other interrupts, or other factors. However, there is still a risk, particularly with regard to the weather.
If a continuous stare is interrupt (for example, due to weather) then the field should be re-acquired
as soon as possible so that the computational demands of shift-and-stack across hours or even nights are
minimized.

\subsection{Estimated time requirement}
\begin{footnotesize}
{\it Approximate total time requested for these observations, using the guidelines available at \url{https://github.com/lsst-pst/survey_strategy_wp}.}
\end{footnotesize}

\vspace{1ex}

\noindent The estimated time required to meet the minimum science goal is around 10.5~hours (2.1~hours for each of five
independent fields).
The estimated time required to meet the additional science goals is around 42~hours (2.1~hours for each
of four visits to each of five independent fields).

\vspace{.3in}

\begin{table}[ht]
    \centering
    \begin{tabular}{|l|l|l|l}
        \toprule
        Properties & Importance \hspace{.3in} \\
        \midrule
        Image quality & 2   \\
        Sky brightness & 1 \\
        Individual image depth & 3  \\
        Co-added image depth & 1  \\
        Number of exposures in a visit   & 3  \\
        Number of visits (in a night)  & 3  \\ 
        Total number of visits & 1  \\
        Time between visits (in a night) & 1 \\
        Time between visits (between nights)  & 1  \\
        Long-term gaps between visits & 1 \\
        Other (please add other constraints as needed) & 3 \\
        \bottomrule
    \end{tabular}
    \caption{{\bf Constraint Rankings:} Summary of the relative importance of various survey strategy constraints. Please rank the importance of each of these considerations, from 1=very important, 2=somewhat important, 3=not important. If a given constraint depends on other parameters in the table, but these other parameters are not important in themselves, please only mark the final constraint as important. For example, individual image depth depends on image quality, sky brightness, and number of exposures in a visit; if your science depends on the individual image depth but not directly on the other parameters, individual image depth would be `1' and the other parameters could be marked as `3', giving us the most flexibility when determining the composition of a visit, for example.}
        \label{tab:obs_constraints}
\end{table}

\subsection{Technical trades}
\begin{footnotesize}
{\it To aid in attempts to combine this proposed survey modification with others, please address the following questions:
\begin{enumerate}
    \item What is the effect of a trade-off between your requested survey footprint (area) and requested co-added depth or number of visits?
    \item If not requesting a specific timing of visits, what is the effect of a trade-off between the uniformity of observations and the frequency of observations in time? e.g. a `rolling cadence' increases the frequency of visits during a short time period at the cost of fewer visits the rest of the time, making the overall sampling less uniform.
    \item What is the effect of a trade-off on the exposure time and number of visits (e.g. increasing the individual image depth but decreasing the overall number of visits)?
    \item What is the effect of a trade-off between uniformity in number of visits and co-added depth? Is there any benefit to real-time exposure time optimization to obtain nearly constant single-visit limiting depth?
    \item Are there any other potential trade-offs to consider when attempting to balance this proposal with others which may have similar but slightly different requests?
\end{enumerate}}
\end{footnotesize}

\vspace{1ex}

\noindent (1) For this program, area cannot be traded against depth, because the science goals require observing faint objects. Deep
(stacked) images are required for this science. (2) ``Rolling cadence''-like observations are not appropriate for this program.
Each visit must be a continuous stare of 2.1~hours so that shift-and-stack can be carried out.
(3) Agnostic, as long as the total stacked image depth of $r=27.5$ (5$\sigma$) is met.
(4) Not relevant for this program. The total stacked image depth is the key requirement.
(5) The primary trade is in exposure time, as described above. 

\section{Performance Evaluation}
\begin{footnotesize}
{\it Please describe how to evaluate the performance of a given survey in achieving your desired
science goals, ideally as a heuristic tied directly to the observing strategy (e.g. number of visits obtained
within a window of time with a specified set of filters) with a clear link to the resulting effect on science.
More complex metrics which more directly evaluate science output (e.g. number of eclipsing binaries successfully
identified as a result of a given survey) are also encouraged, preferably as a secondary metric.
If possible, provide threshold values for these metrics at which point your proposed science would be unsuccessful 
and where it reaches an ideal goal, or explain why this is not possible to quantify. While not necessary, 
if you have already transformed this into a MAF metric, please add a link to the code (or a PR to 
\href{https://github.com/lsst-nonproject/sims_maf_contrib}{sims\_maf\_contrib}) in addition to the text description. (Limit: 2 pages).}
\end{footnotesize}

\vspace{1ex}

\noindent The key science requirement (metric) is the uncertainty on the number of faint KBOs. With a sample size of 
$10^4$~objects, as described above, this uncertainty naively is something like 1\%. If the uncertainty were to
approach 10\% or more than the power of this science investigation is dramatically reduced. However, an
uncertainty of 2\% (for example) does not weaken the science case significantly from the nominal case. 
In the case of such a trade, the increased uncertainty should be introduced by including fewer pointings (less area)
and 
not by shallower survey fields, which would not allow probes of the smallest KBOs.

\vspace{1ex}

\noindent The requirements for the enhanced science goals are twofold. In the first case, the metric is the number of objects
observed at $>$30$\sigma$ in two different colors (to allow for color diagnostic science with color
uncertainties on the order of 0.05~mag). The number of objects should be at least
100~in each of the major dynamical classes --- implying some 1000~objects total --- so that 
each sub-class of KBOs is well sampled
in order to detect what is expected to be 
a relatively subtle signature. In the second case, the metric is the orbit quality (which leads to dynamical classification).
Arc length is an acceptable proxy for orbit quality. Here the requirement is around three years, and includes 
at least one non-opposition measurement (for additional orbit-fitting power).

\vspace{1ex}

\noindent Thus, the metrics should be the following:

\begin{description}
\item{[1]} Fracitonal uncertainty (as root(N)) on the number of faint ($r=27.5$) KBOs. 1\% is the science goal, and $<$10\% is acceptable.
\item{[2]} Number of objects observed at $>$30$\sigma$ in both bands; at least 100~objects in each
dynamical subclass, or $\sim$1000 objects total.
\item{[3]} Arc length for observed objects. Two years is the goal. One year arcs carry significantly less
diagnostic power.
\end{description}

\vspace{1ex}

\noindent It would be appropriate to use Lynne Jones' KBO population in {\tt opsim}, for example, to test the
impact of various cadences on the science output described here.

\vspace{.6in}

\section{Special Data Processing}
\begin{footnotesize}
{\it Describe any data processing requirements beyond the standard LSST Data Management pipelines and how these will be achieved.}
\end{footnotesize}

\vspace{1ex}

\noindent KBOs at 40~AU move at 
around 3''/hour. 
To detect ultra-faint KBOs, ideally one would track not at the sidereal rate but at a rate that keeps the target object stationary. However, there are many such objects in an LSST field, and their rates and directions of motion vary,
are not known {\em a priori}.
The only successful observing technique is to use exposures shorter than 10~minutes so that
trailing is negligible.

\vspace{1ex}
					
\noindent The solution is to take a sequence of shorter exposures, and perform an offline ``shift-and-stack'' procedure in a grid of position angle and velocity, a method first applied to KBO searches by Bernstein [8].
The standard LSST imaging processing pipeline will produce reduced data for individual frames. This standard procedure
is then followed by ``shift-and-stack'' or ``digitally tracking'' analysis.
In these ``digitally tracked'' images, KBOs tracked at the correct rate will appear as point sources
(sidereal sources are masked before stacking).
The success of this technique is shown in Figure~\ref{stack}: using $11\times 330$s $z$-band sequences of the DECam/Dark 
Energy Survey X3 deep supernova search field a $16\sigma$ detection of the 24.4-mag extreme KBO 2013~RF$_{98}$ ($a=350$~AU) was recovered.
Each of the 2.1~hour visits 
will consist of a series of exposures (30~or 600~seconds, with this decision to be studied as a trade, as described
above).
For an exposure sequence of this duration, it is sufficient to stack in a grid of position angle and velocity; higher-order contributions from track curvature are negligible.

\section{Acknowledgements}
 \begin{footnotesize}
 {\it If you have any special acknowledgements of support for the preparation of this white paper, please feel free to use this section. If not, feel free to comment out.}
 \end{footnotesize}

\vspace{1ex}

\noindent This White Paper benefitted from the Solar System Science Collaboration Sprint, sponsored by LSSTC and B612, that was
held in Seattle in July, 2018. We acknowledge important early work on
Solar System DDFs by Becker et al.\footnote{{\tt http://lsst-sssc.github.io/Files/Becker-solarsystem-01.pdf}}.

\section{References}

[1] Brown 2012, AREPS, 40, 467.
[2] Schlichting et al.\ 2013, AJ, 146, 36.
[3] Schlichting et al.\ 2012, ApJ, 761, 150.
[4] Chang et al.\ 2013, MNRAS, 429, 1626.
[5] Liu et al.\ 2015, MNRAS, 446, 932.
[6] Shankman et al.\ 2013, ApJL, 764, L2.
[7] Belton 2014, Icarus, 231, 168.
[8] Bernstein et al.\ 2004, AJ, 128, 1364.
[9] Fuentes et al.\ 2010, ApJ, 722, 1290.
[10] Fuentes et al.\ 2012, ApJ, 742, 118.
[11] Parker et al.\ 2015, LPSC, 46, 2614.
[12] Trilling 2016, PLoS, e0147386.
[13] McKinnon et al.\ 2016, Nature, 534, 82.
[14] Bannister et al.\ 2016, AJj, 152, 70.
[15] Schwamb et al.\ 2018, ApJS, submitted (arXiv: 1809.08501)

\end{document}